# The SAFARI Detector System

Michael D. Audley*[a], Gert de Lange[a], Jian-Rong Gao[b,c], Brian D. Jackson[a]
Richard A. Hijmering[b], Marcel L. Ridder[b], Marcel P. Bruijn[b], Peter R. Roelfsema[a], Peter A. R. Ade[d],
Stafford Withington[e], Charles M. Bradford[f], Neal A. Trappe[g]

[a]SRON Netherlands Institute for Space Research, Postbus 800, 9700 AV Groningen, The Netherlands; [b]SRON Netherlands Institute for Space Research, Sorbonnelaan 2, 3584 CA Utrecht, The Netherlands; [c]Kavli Institute of Nanoscience, Delft University of Technology, Lorentzweg 1, 2628 CJ Delft, The Netherlands; [d]School of Physics and Astronomy, Cardiff University, 5 The Parade, Cardiff CF24 3AA, UK; [e]Cavendish Laboratory, University of Cambridge, JJ Thomson Avenue, Cambridge CB3 0HE, UK; [f]Jet Propulsion Laboratory, M/S 169-506, 4800 Oak Grove Drive, Pasadena, CA 91109, USA; [g]Department of Experimental Physics, Maynooth University, Maynooth, Co. Kildare, Ireland

**ABSTRACT**

We give an overview of the baseline detector system for SAFARI, the prime focal-plane instrument on board the proposed space infrared observatory, SPICA. SAFARI's detectors are based on superconducting Transition Edge Sensors (TES) to provide the extreme sensitivity (dark NEP$\leq 2\times 10^{-19}$ W/$\sqrt{Hz}$) needed to take advantage of SPICA's cold (<8 K) telescope. In order to read out the total of ~3500 detectors we use frequency domain multiplexing (FDM) with baseband feedback. In each multiplexing channel, a two-stage SQUID preamplifier reads out 160 detectors. We describe the detector system and discuss some of the considerations that informed its design.

## 1. INTRODUCTION

The infrared satellite observatory SPICA[1] has been selected as one of three candidate missions for the European Space Agency's M5 program. SPICA's telescope will use a large (2.5-m diameter) primary mirror cooled to $\leq 8$ K to enable high spectral-resolution, sky-background limited observations of the cold dusty Universe in the mid- and far-infrared. SPICA's complement of focal-plane instruments comprises a mid-infrared spectrometer/camera (SMI), a polarimetric camera (POL), and a far-infrared spectrometer (SAFARI). These three instruments will combine to give SPICA a total spectral range of $\lambda$=12—230 µm. The mission promises to revolutionize our knowledge of the origin and evolution of galaxies, stars and planetary systems[2],[3],[4],[5],[6],[7]. The SAFARI instrument was originally conceived as a far-infrared imaging Fourier transform spectrometer (FTS) for the SPICA satellite[8],[9],[10]. Since then, the design has evolved through detailed studies and has been optimized for ultra-sensitive point-source spectroscopy while retaining some mapping capability[11]. SAFARI now uses diffraction gratings to provide a default spectral resolution of ~300 and an in-line Martin-Puplett FTS[12] for a high spectral-resolution mode with a resolution varying with wavelength from ~1500 at 230 µm to ~11000 at 34 µm. SAFARI will cover the spectral range $\lambda$=34—230 µm simultaneously with four grating modules, each containing a diffraction grating and a 3×294-element detector array to sample the dispersed spectrum. The four detector arrays will contain a total of ~3500 Transition Edge Sensor[13] (TES) bolometers using superconducting bilayers on thin, thermally isolated silicon nitride islands[14],[15],[16]. The bolometers sit behind few-moded feedhorns and in front of reflecting backshorts[17]; incoming radiation is absorbed by a resistive film.

*audley@physics.org; phone +31 (0)50 363 9361; fax +31 (0)50 363 4033; www.sron.nl

SAFARI will allow the studies of the dynamics and chemistry of a wide range of objects, including galaxies out to redshifts of $z \approx 5$—6. To take advantage of SPICA's low-background cold mirror, SAFARI's detectors must achieve a dark noise equivalent power (NEP) of $2 \times 10^{-19}$ W/√Hz. To achieve the reduction in thermal noise to reach this sensitivity, the TES detectors have a transition temperature, $T_c$, of about 100 mK and are operated with a bath temperature of 50 mK. The dispersion of the grating reduces the optical power, and hence the photon noise on each detector. Detectors with an NEP of $2 \times 10^{-19}$ W/√Hz lead to a line sensitivity in the low-resolution (R=300) mode of $\sim 5 \times 10^{-20}$ W/m$^2$ for a 5σ detection in a 1-hour observation[1]. The instrument is detector-noise limited, so any future improvement in detector NEP could improve SAFARI's sensitivity even further.

## 2. THE SAFARI INSTRUMENT

### 2.1 Overview

The SAFARI instrument comprises four grating modules, each providing the dispersion for the default, R=300 mode. The four grating modules cover the following wavelength ranges: SW: 34–56 μm, MW: 54–89 μm, LW=87–143 μm, and VLW: 140–230 μm. A Martin-Puplett interferometer will be used inline to provide a high resolution mode. A two-dimensional beam steering mirror (BSM) in an Offner relay is used to direct radiation from the sky or from an internal calibration source either directly to the low-resolution grating spectrometer or via the Martin-Puplett interferometer. The BSM is also used for source/background chopping and for implementing small-area (<2') mapping observations. The optical design is described in detail elsewhere[18].

### 2.2 Grating Modules

The grating module is enclosed in a light-tight box with an entrance aperture that is designed to illuminate the grating. A detailed block diagram showing the internals of the grating module is shown in Figure 1. The nominal temperature of the enclosure is 1.8 K. This enclosure contains the grating-spectrometer optics as well as the detectors and cold readout electronics. Inside the 1.8K enclosure is a 300mK thermal shield in which is mounted the 50mK focal plane assembly. The almost-closed geometry of the 1.8K enclosure lends itself naturally to magnetic shielding. The baseline is to have a magnetic shield around the grating module. This baseline architecture has a shield around the optical components of the grating module. Since these optical components do not need magnetic shielding, other shielding architectures, such as locally shielding the 50mK focal plane assembly would be attractive.

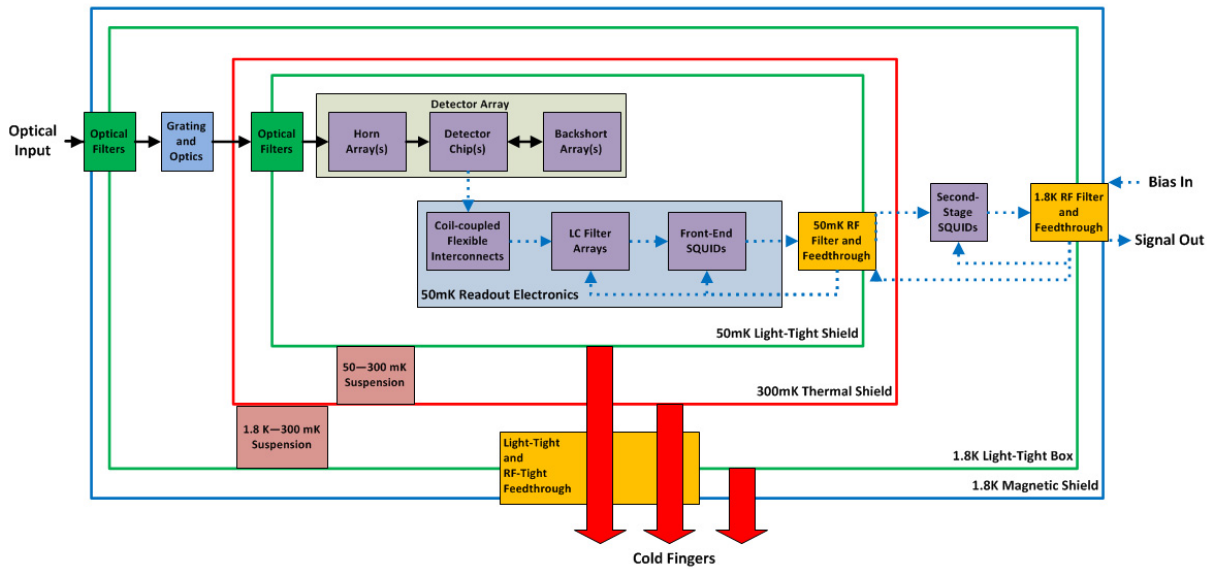

Figure 1. Block diagram of a grating module.

## 3. FOCAL-PLANE ASSEMBLY

The focal-plane assembly contains the detectors, feedhorns, and cold readout electronics in a light-tight enclosure, with a band-pass filter in front of the feedhorns. In this section we describe the detectors and the readout, and conclude with a description of how the detectors are laid out in the focal-plane assembly.

### 3.1 Detectors

Figure 2 shows a TES bolometer fabricated at SRON[14]. SAFARI's TES bolometers are two orders of magnitude more sensitive than bolometers used for ground-based observations[19] and testing them requires careful attention to stray-light protection and shielding from electrical, magnetic, and mechanical disturbances[20]. The detector requirements are within reach: we have demonstrated the required NEP$\leq 2\times 10^{-19}$ W/$\sqrt{Hz}$[21] and optical efficiency $\geq$50% for single detectors[22]. More recent optical measurements of arrays fabricated at the University of Cambridge are described elsewhere in these proceedings[23].

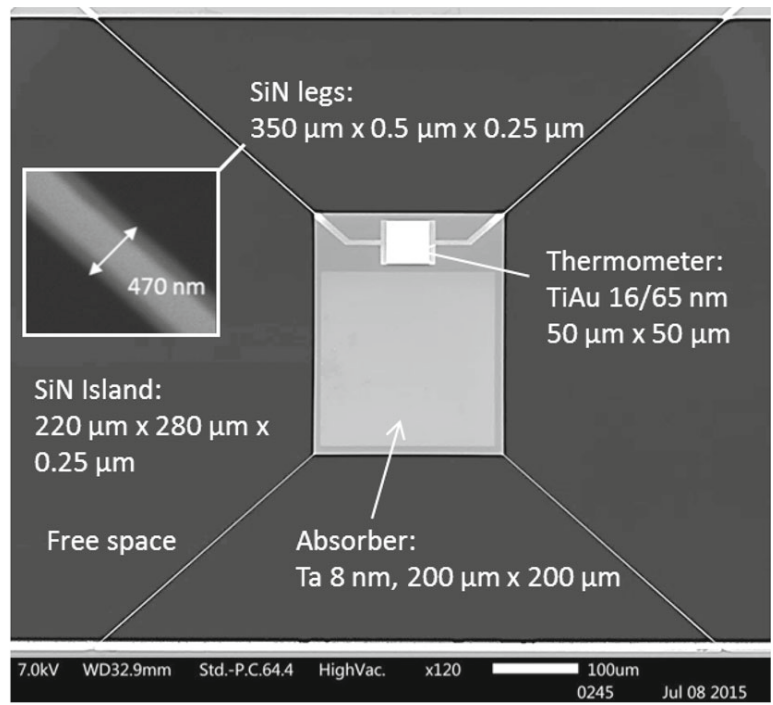

Figure 2. A SEM image of a prototype TES bolometer for SAFARI[14].

### 3.2 Readout

In order to reduce the wire count, and hence the thermal load, the ~3500 detectors are read out using frequency-domain multiplexing with baseband feedback[24],[25]. Each TES bolometer is AC-biased, at a unique frequency between ~1 and ~4 MHz. A narrow-band LC filter[26] in series with each TES ensures that the TES sees only its assigned bias signal. The LC filters also reject wide-band Johnson noise in the circuit. All the TES bias signals can be adjusted independently so that all detectors are biased optimally. The TES currents are summed and read out with a SQUID at 50 mK (the front-end SQUID). A second SQUID amplifier stage on the 1.8K temperature level of the grating module provides additional amplification. The SQUID output signal is essentially an amplitude-modulated version of the AC bias signal, with a phase shift due to the electronics. After further amplification by a low noise amplifier at a higher temperature, the signal is demodulated and a comb of frequencies is generated with the appropriate phases and amplitudes to cancel out the SQUID output when applied to the feedback coil of the front-end SQUID. This feedback signal is the quantity that is linearly proportional to the power absorbed by each TES bolometer. Because the SQUID output is essentially an amplitude-modulated version of the AC bias, the room temperature electronics just need to generate an amplitude-modulated version of the AC-bias signal, i.e. the feedback is carried out on the amplitude of the generated tones. This amplitude varies no faster than the bandwidth of the TES, not at the ~MHz frequencies of the AC bias. Thus, this scheme is known as baseband feedback. In SAFARI's baseline design, one channel will read out 160 detectors. This has been demonstrated to be feasible[27].

### 3.3 Focal-Plane Layout

The layout of the focal plane is shown in Figure 3. There are three rows of detectors. Each row corresponds to a single spatial pixel on the sky, with the three spatial pixels in a line on the sky spanning SAFARI's field of view (2'). The 294 spectral pixels in each row fully sample the 147 elements of the dispersed spectrum. Adjacent rows of detectors are offset by one third of a spectral pixel. This will ensure that a narrow (unresolved) line that falls on a spectral-pixel boundary in one of the spatial pixels will be measured well, even if it falls on a dead spectral pixel in one of the rows. The spacing between rows of detectors, i.e. between spatial pixels, is $\Delta x = 4F_2\lambda$ to ensure sufficient separation to prevent contamination between on-source and off-source pointings for background subtraction. Compared with the square arrays of SAFARI's original design concept, this layout makes detector packing and wiring easier, due to the increased space available in the spatial direction.

We chose the number of spatial pixels, i.e. detector rows, to be three, rather than two, in order to avoid gaps in spectral coverage due to possible detector failures. With more than one spatial pixel, when we chop on and off source for background subtraction, we can ensure there is always a spatial pixel on source, so that there is no observing-time penalty for background subtraction.

The beam from the grating spectrometer falling on the detectors is anamorphic, i.e. the focal ratio is different in the spectral ($F_1$) and spatial ($F_2$) directions. The baseline spectral pixel size is $1.5F_1\lambda$. We chose a pixel spacing in the spectral direction of 800 μm. This choice was driven by the minimum detector size for the required NEP and speed and the minimum feedhorn wall thickness (100 μm). Also, a smaller spectral-pixel pitch in the VLW grating module would have led to extremely small spectral-direction focal ratio ($F_1$). The physical size of the focal plane is set by the focal ratio and wavelength and is slightly different for each grating module, ranging from 7×239 mm for the SW band to 9×251 mm for the VLW band.

The total number of detectors is limited by the thermal conductance of the readout wiring. In the baseline design, we have a total of 3840 detectors for the four grating modules. With a multiplexing factor of 160, these could be read out with 24 readout channels, with enough excess readout capacity for up to 10 blind pixels per channel.

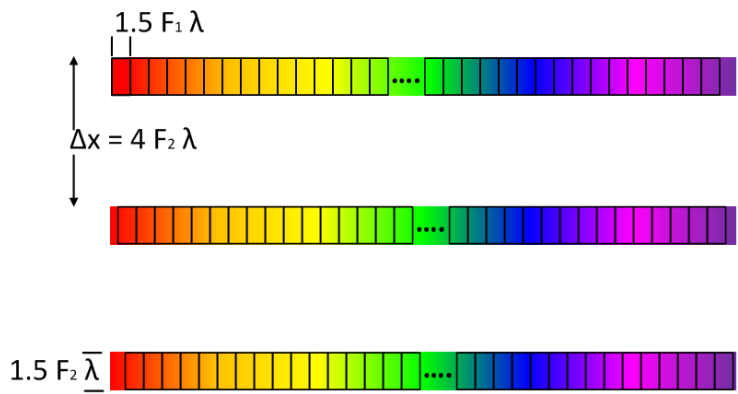

Figure 3. Detector array layout. Each rectangle corresponds to the aperture of a feedhorn, color-coded to show the dispersion direction. Each row of detectors corresponds to a single spatial pixel on the sky. Adjacent rows are offset by one third of a spectral pixel. The central pixels of each row have been omitted for clarity: there will actually be 294 spectral pixels in each spatial pixel.

Given the length of the focal plane (~250 mm), it would be impractical to fabricate monolithic focal plane arrays. Therefore, each focal plane array is made up of shorter sub-array modules. A possible configuration has six sub-array modules, each containing an array of 3×49 detectors, making up each of the four 3×294 focal-plane arrays. Then a sub-array module will approximately match a 160-detector readout channel. For redundancy, it might be desirable to cross-strap pairs of sub-array modules to two readout channels. This would prevent any loss of spectral coverage if a readout channel fails.

## 4. SUMMARY

We have described the SAFARI detector system and explained some of the reasoning that went into its design. It should be noted that the design presented here is the current baseline. Detailed studies are under way to optimize the design. Some of the design parameters quoted here may change as the design is refined in order to best satisfy SAFARI's science requirements. The features and performance of SAFARI are summarized in Table 1. The sensitivity values here are derived from a detector NEP of $2\times10^{-19}$ W/$\sqrt{Hz}$; any future improvement in detector NEP could improve SAFARI's sensitivity even further.

Table 1. Features of SAFARI's four wavelength bands (SW, MW, LW, and VLW) and performance of SAFARI[1] in its low-resolution (LR) and high-resolution (HR) modes. Note that the design parameters may change as the design is optimized.

| Instrument Features: | SW | | MW | | LW | | VLW | |
|---|---|---|---|---|---|---|---|---|
| Wavelength Range (μm) | 34–56 | | 54–89 | | 87–143 | | 140–230 | |
| Band Center (μm) | 45 | | 72 | | 115 | | 185 | |
| FWHM @ Band Center (arcsec) | 4.5 | | 7.2 | | 12 | | 19 | |
| Number of Spatial / Spectral Pixels | 3/294 | | 3/294 | | 3/294 | | 3/294 | |
| Spectral resolution in low-resolution (LR) mode | 300 | | 300 | | 300 | | 300 | |
| Spectral resolution in high-resolution (HR) mode | 11700–7150 | | 7400–4500 | | 4600–2800 | | 2850–1740 | |
| Pixel Size (spatial×spectral; μm) | 1400×802 | | 1200×801 | | 1220×802 | | 1540×800 | |
| Focal Plane Size (mm) | 7×239 | | 7×246 | | 8×247 | | 9×251 | |
| Instrument Performance: | LR | HR | LR | HR | LR | HR | LR | HR |
| Point-source Spectroscopy (5σ-1 hr) limiting flux ($10^{-20}$ W/m$^2$) | 7.2 | 13 | 6.6 | 13 | 6.6 | 13 | 8.2 | 15 |
| Point-source Spectroscopy (5σ-1 hr) limiting flux density (mJy) | 0.31 | 18 | 0.45 | 17 | 0.72 | 17 | 1.44 | 19 |
| Mapping Spectroscopy 1'×1' (5σ-1 hr) limiting flux ($10^{-20}$ W/m$^2$) | 84 | 189 | 49 | 113 | 30 | 73 | 23 | 51 |
| Mapping Spectroscopy 1'×1' (5σ-1 hr) limiting flux density (mJy) | 3.6 | 253 | 3.3 | 151 | 3.3 | 97 | 4.1 | 67 |
| Photometric Mapping 1'×1' (5σ-1 hr) limiting flux density (mJy) | 209 | | 192 | | 194 | | 239 | |
| 5σ Confusion Limit (mJy) | 15 | | 200 | | 2000 | | 10000 | |